\documentclass[vecphys]{svmult}

\usepackage{makeidx}

\usepackage{graphicx}

\usepackage{multicol}

\usepackage[bottom]{footmisc}

\makeindex

\begin{document}

\title*{Vector Boson Pair Production via Vector Boson Fusion at NLO QCD \thanks{Presented by G.~Bozzi at IFAE 2007 (Napoli, April 2007) and HEP 2007 (Manchester, July 2007)}}

\author{Giuseppe Bozzi\inst{1}, Barbara J\"ager\inst{2}, Carlo Oleari\inst{3} \and Dieter Zeppenfeld\inst{1}}

\institute{Institut f\"ur Theoretische Physik, Universit\"at Karlsruhe, P.O.Box 6980, 76128 Karlsruhe, Germany
\and KEK Theory Division, Tsukuba 305-0801, Japan
\and Universit\`a di Milano Bicocca and INFN Sezione di Milano Bicocca, 20126 Milano, Italy}

\maketitle

\textbf{Abstract.} NLO QCD corrections to Vector Boson Pair Production via Vector Boson Fusion have recently been calculated and implemented in a parton-level Monte-Carlo program with full experimental cuts. We briefly sketch the elements of the calculation and show numerical results for the Large Hadron Collider.

\subsubsection*{Introduction}\label{intro}

\vspace{-0.25cm}

The Vector Boson Fusion (VBF) process $qq \to qqH$ is one of the most promising channels for the discovery of the Higgs particle and the measurement of its properties at the Large Hadron Collider (LHC) \cite{VBF:C}. It proceeds through a $t-$channel scattering of the two initial-state quarks mediated by a weak boson, with the Higgs emitted off the boson propagator. The kinematic features that make this process phenomenologically relevant are: the presence of two highly-energetic jets in the final state, the large rapidity interval between these jets and the absence of noticeable jet activity inside this rapidity interval.

Moreover the Next-to-Leading Order (NLO) QCD corrections to the total cross
sections \cite{Han:1992hr} and for the differential distributions
\cite{Figy:2003nv} have been found to be quite modest (enhancing the Leading
Order (LO) result by 5-10\%), thus pointing towards a good stability of the perturbative result.

Even though the cross section for the VBF process is somewhat smaller than the one for the gluon fusion $gg \to H$ channel at the LHC, the distinctive features cited above greatly help in distinguishing the signal from the backgrounds and make VBF an ideal candidate for Higgs discovery and precision measurements.

One of the most relevant backgrounds to a VBF $H\to VV$ signal is the process $qq \to qqVV$, i.e. vector boson pair production via VBF \cite{VBF:H}. It shows exactly the same kinematical features as VBF Higgs production, thus being an {\it irreducible} background. In addition, it is known that the scattering of longitudinal vector bosons is intimately related to the mechanism of electroweak symmetry breaking (EWSB), and an enhancement of $qq \to qqVV$ over the Standard Model predictions at high center-of-mass energies could be a possible signal of {\it strong} EWSB (see, for instance, \cite{Chanowitz:2004gk} and references therein).

It is thus clear that an accurate prediction for the electroweak production of a vector boson pair plus two jets is mandatory for new physics searches at the LHC.

In the following we will present the results obtained in three recent papers where we computed the NLO QCD corrections to the processes $qq \to qqW^+W^-$ \cite{Jager:2006zc}, $qq \to qqZZ$ \cite{Jager:2006cp}, $qq \to qqW^{\pm}Z$ \cite{Bozzi:2007ur}, including the full leptonic decays of the vector bosons. The calculations have been implemented in a fully-flexible parton level Monte-Carlo program, allowing for the computation of jet observables and a straightforward implementation of experimental cuts.

\subsubsection*{Selected topics from the calculation}\label{topics}

\vspace{-0.25cm}

The main challenges of the calculation were the huge number of Feynman diagrams involved and the numerical instabilities arising from pentagon contributions to the virtual part of the cross section.

When computing a multi-parton process (in this case, a 2$\to$6 process considering full leptonic decays of the vector bosons) one has to find an efficient way to speed up the numerical evaluation. In our Monte-Carlo code the sum of sub-amplitudes encountered in several diagrams, and involving only EW bosons and leptons, is computed only once per phase-space point (for details, see \cite{Jager:2006zc}): this method is particularly efficient in the computation of real emission corrections, due to the large number of contributing diagrams.

The soft and collinear singularities of the NLO real contributions have been taken into account by means of the standard Catani-Seymour dipole subtraction \cite{Catani:1996vz}. Since the divergences only depend on the colour structure of the external partons, the subtraction terms are {\it identical} in form to the ones appearing in Higgs production via VBF \cite{Figy:2003nv}.

The EW bosons exchanged in the $t-$channel are colour-singlets, thus there
cannot be any virtual contribution at ${\cal O (\alpha_S)}$ from gluons
attached both to the upper and lower quark lines. This allows us to consider
virtual radiative corrections {\it separately} for the single quark lines,
leading to diagrams containing loops with up to five external legs ({\it
  pentagons}). After the cancellation of the infrared poles between real and
virtual contributions, the finite remainder of the loop amplitudes can be
computed by means of the Passarino-Veltman reduction formalism
\cite{Passarino:1978jh} in the case of two-, three- and four-point tensor
integrals. In the case of pentagons, numerical instabilities show up when
kinematical invariants, such as the Gram determinant, become small for some
regions of phase space. For the pentagons, we have thus used the recently
proposed alternative reduction formalism by Denner and
Dittmaier~\cite{Denner:2002ii}, which results in a fraction of
numerically unstable events at the per-mille level (for details, see \cite{Bozzi:2007ur}).

\subsubsection*{Numerical results}\label{numres}

\vspace{-0.25cm}

In the following we will present numerical results at NLO QCD accuracy obtained with our Monte-Carlo code for EW $^+W^-Wjj, ZZjj, W^{\pm}Zjj$ production at the LHC.

We used the CTEQ6M parton distributions with $\alpha_S$=0.118 at NLO and the CTEQ6L1 set at LO \cite{Pumplin:2002vw}. We chose $m_Z$=91.188 GeV, $m_W$=80.419 GeV and $G_F$=1.166$\times 10^{-5}$ as EW input parameters, obtaining $\alpha_{QED}$=1/132.54 and $\sin^2\theta_W$=0.22217. We have set fermion masses to zero, neglecting external bottom and top quark contributions. Jets have been reconstructed by means of the $k_T$-algorithm \cite{kToriginal,kTrunII} with resolution parameter $D$=0.8.

Typical VBF cuts have been imposed: here we show those used in the $W^{\pm}Z$ case:

\begin{itemize}

\item two hard ''tagging'' jets:\,\, $p_{Tj}\geq$20 GeV,\,\, 

$|y_j|$$\leq$4.5,\,\, $M_{jj}>600$~GeV

\item large rapidity separation between jets:\,\, $\Delta y_{jj}>$4,\,\, $y_{j1}\times y_{j2}<$0

\item lepton cuts:\,\, $p_{Tl}\geq$20 GeV,\,\, $|\eta_l|\leq$2.5,\,\, $m_{ll}\geq$15 GeV

\item ''separation'' cuts: $\Delta R_{jl}\geq$0.4,\,\, $\Delta R_{ll}\geq$0.2\,,

\end{itemize}

where $\Delta R_{jl}$ and $\Delta R_{ll}$ denote the jet-lepton and lepton-lepton separation in the rapidity-azimuthal angle plane, respectively, and $m_{ll}$ the invariant mass of an electron or muon pair.

\begin{figure}

\centering

\includegraphics[scale=0.7]{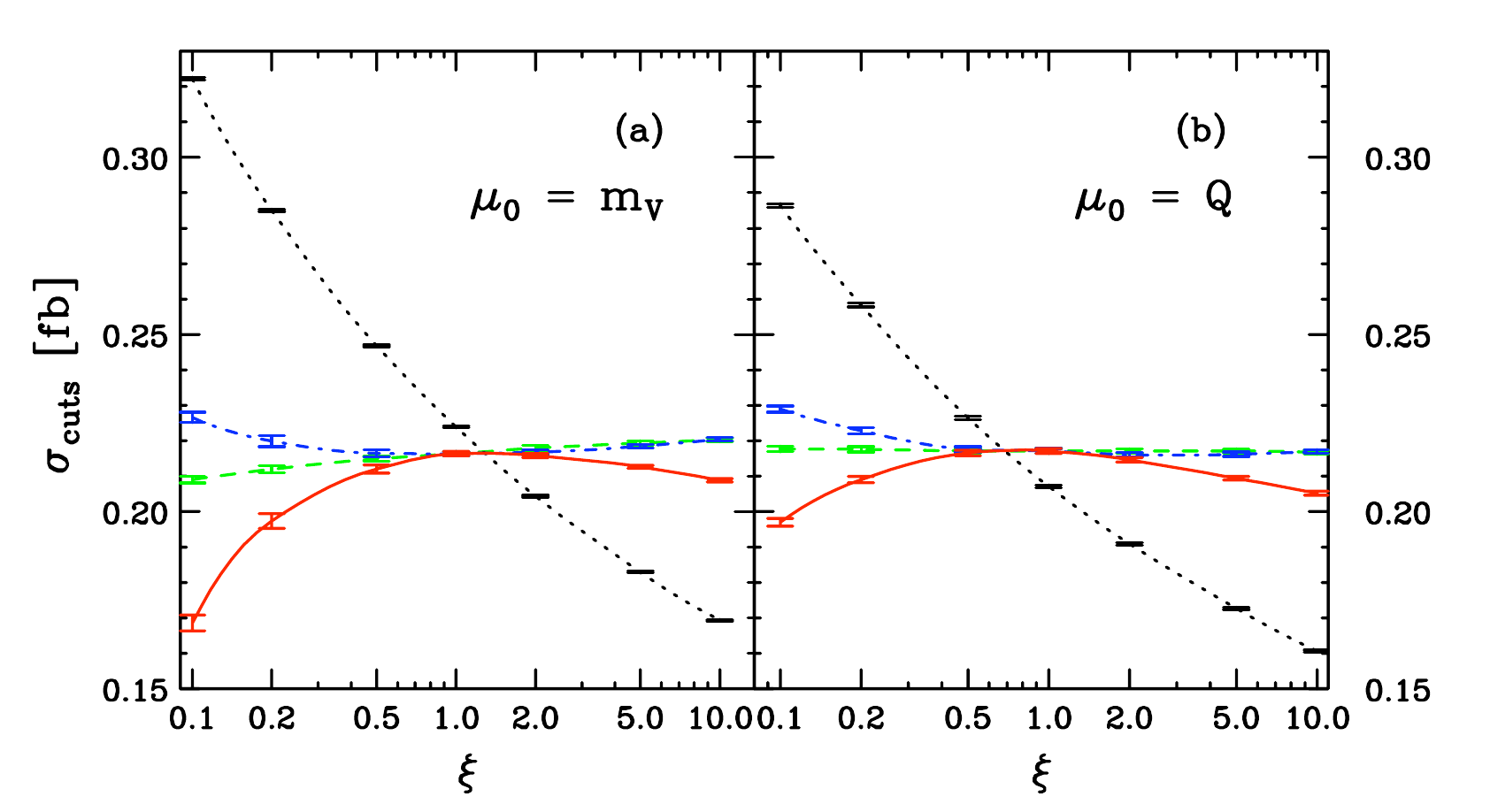}

\caption{Scale dependence of the total cross section for $W^{\pm}Z$ production via VBF for two values of the central scale (see text).}

\label{fig:1}

\end{figure}

In Figure \ref{fig:1} (from \cite{Bozzi:2007ur})  we show the total cross
section displayed as a function of the factorization and renormalization
scales $\mu_{F,R}=\xi_{F,R}\mu_0$ in the case of $W^{\pm}Zjj$ production. We
have considered two possible values for the central scale: $\mu_0=(m_Z+m_W)/2$
or $\mu_0=Q$, where $Q$ is the momentum transfer carried by the exchanged
vector boson in VBF graphs (for details, see \cite{Bozzi:2007ur}). The
K-factor is $K$=0.97 in the first case and $K$=1.04 in the second one. In both
cases the scale dependence, for instance in the range $0.5<\xi<2$, is
greatly reduced when passing from LO (about 10\%, dotted black curves) to NLO
(about 2\%). At NLO, we have considered three cases: $\xi_F=\xi_R=\xi$ (solid red lines), $\xi_F=1, \xi_R=\xi$ (dashed green lines), $\xi_F=\xi, \xi_R=1$ (dot-dashed blue lines).

\begin{figure}

\centering

\includegraphics[scale=0.6]{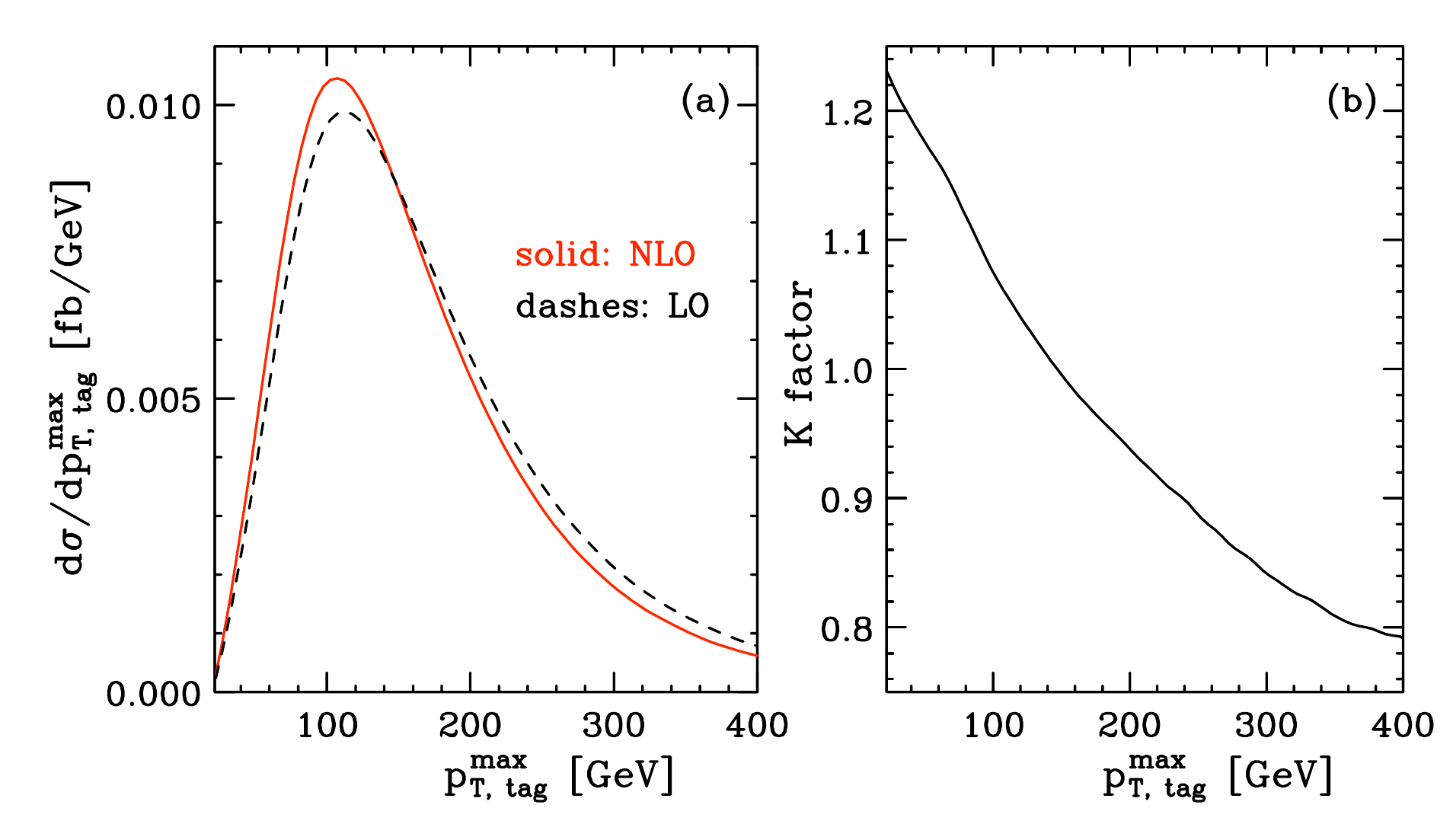}

\caption{Transverse-momentum distribution of the tagging jet with the highest
  $p_T$ in $W^+W^-$ production via VBF (left), with the corresponding
  $K$-factor (right). The cuts of Ref.~\cite{Jager:2006zc} are used.}

\label{fig:2}

\end{figure}

In Figure \ref{fig:2} (from \cite{Jager:2006zc}) we present the transverse-momentum distribution of the tagging jet with the highest $p_T$ in the $WWjj$ case, together with the corresponding K-factor. The figure shows a strong change in shape when going from LO to NLO, with a 10-20\% enhancement of the cross section at low values of $p_T$ ($p_T<$100 GeV) and a corresponding decrease at higher $p_T$ values: this effect is mainly due to the extra parton coming from real emission at NLO.

\begin{figure}

\centering

\includegraphics[scale=0.7]{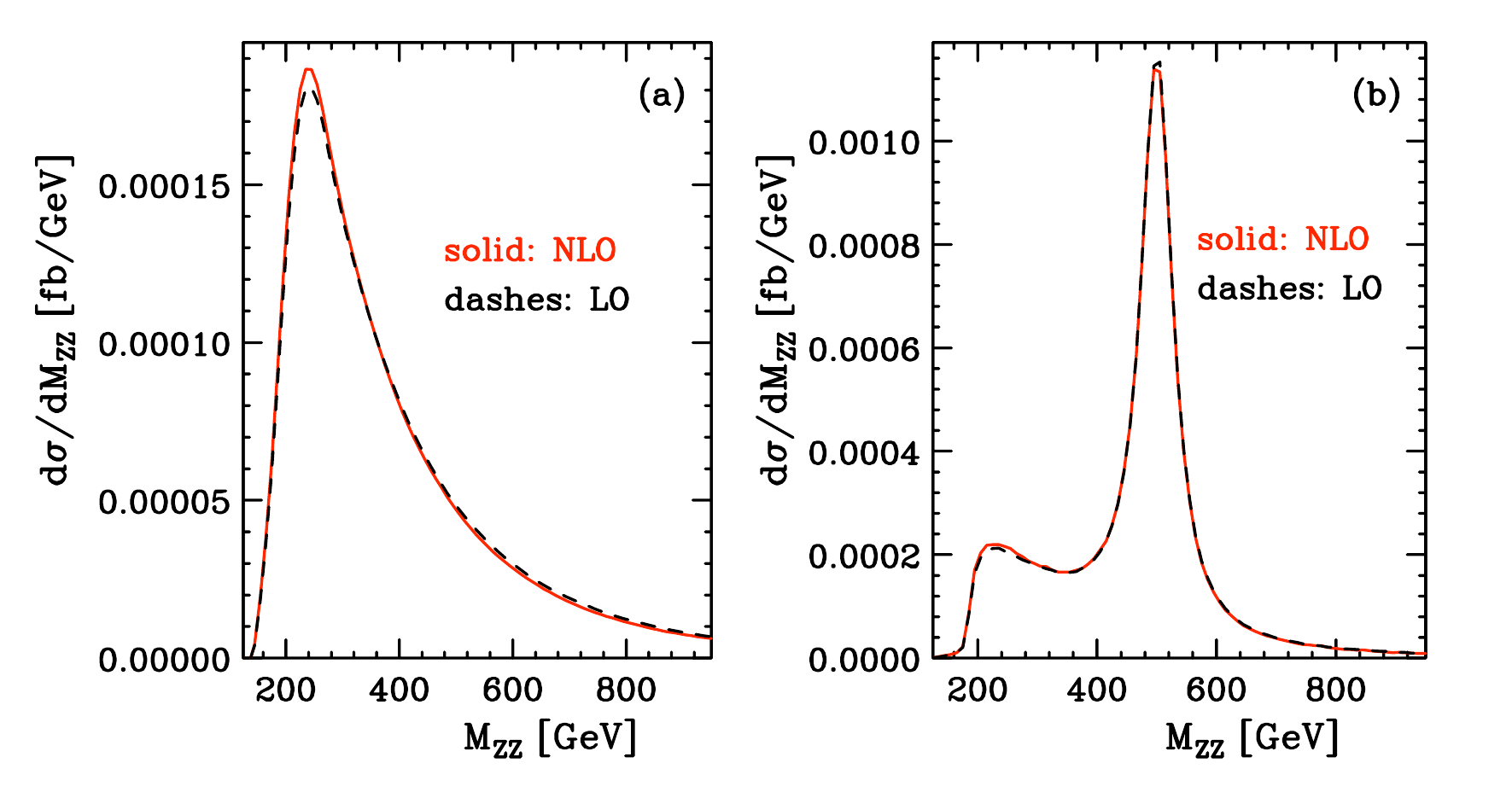}

\caption{Distribution of the invariant mass $M_{ZZ}$ in $ZZ$ production via
  VBF without (left) and with (right) the Higgs boson contribution. The cuts
  of Ref.~\cite{Jager:2006cp} are used.}

\label{fig:3}

\end{figure}

Finally, in Figure \ref{fig:3} (from \cite{Jager:2006cp}) we show the
differential distribution for EW $ZZjj$ production with respect to the
invariant mass $M_{ZZ}$ without (left) and with (right) the inclusion of the
Higgs contribution for a scale $\mu=Q$. Apart from the pronounced resonance
behaviour visible in the right plot, we note that the LO and NLO predictions are virtually indistinguishable in both cases, indicating an excellent stability of the perturbative calculation for this scale choice.

{\bf Acknowledgements.} The work of B.J. is supported by the Japan Society for the Promotion of Science. The work of G.B. is supported by the Deutsche Forschungsgemeinschaft under SFB TR-9 ``Computergest\"utzte Theoretische Teilchenphysik''.

\printindex


\vspace{-0.5cm}



\begin{thebibliography}{99.}

\bibitem{VBF:C}
  D.~Zeppenfeld, R.~Kinnunen, A.~Nikitenko, and E.~Richter-Was,
  Phys.\ Rev.\ D {\bf 62}, 013009 (2000)
  [arXiv:hep-ph/0002036];
  D.~Zeppenfeld,
  in {\it Proc. of the APS/DPF/DPB Summer Study on the Future of
    Particle Physics, Snowmass, 2001} edited by N.~Graf,
  eConf {\bf C010630}, p.\ 123 (2001)
  [arXiv:hep-ph/0203123];
  A.~Belyaev and L.~Reina,
  JHEP {\bf 0208}, 041 (2002)
  [arXiv:hep-ph/0205270];
  M.~D\"uhrssen et al.,
  Phys.\ Rev.\ D {\bf 70}, 113009 (2004)
  [arXiv:hep-ph/0406323].
  
\bibitem{Han:1992hr}
  T.~Han, G.~Valencia and S.~Willenbrock,
  Phys.\ Rev.\ Lett.\  {\bf 69}, 3274 (1992)
  [arXiv:hep-ph/9206246].

\bibitem{Figy:2003nv}
  T.~Figy, C.~Oleari and D.~Zeppenfeld,
  Phys.\ Rev.\  D {\bf 68}, 073005 (2003)
  [arXiv:hep-ph/0306109].

\bibitem{VBF:H}
  D.~Rainwater and D.~Zeppenfeld,
  Phys.\ Rev.\ D {\bf 60}, 113004 (1999)
  [Erratum-ibid.\ D {\bf 61}, 099901 (2000)]
  [arXiv:hep-ph/9906218];
  N.~Kauer, T.~Plehn, D.~Rainwater, and D.~Zeppenfeld,
  Phys.\ Lett.\ B {\bf 503}, 113 (2001)
  [arXiv:hep-ph/0012351].

\bibitem{Chanowitz:2004gk}
  M.~S.~Chanowitz,
  Czech.\ J.\ Phys.\  {\bf 55}, B45 (2005)
  [arXiv:hep-ph/0412203].

\bibitem{Jager:2006zc}
  B.~Jager, C.~Oleari and D.~Zeppenfeld,
  JHEP {\bf 0607}, 015 (2006)
  [arXiv:hep-ph/0603177].

\bibitem{Jager:2006cp}
  B.~Jager, C.~Oleari and D.~Zeppenfeld,
  Phys.\ Rev.\  D {\bf 73}, 113006 (2006)
  [arXiv:hep-ph/0604200].

\bibitem{Bozzi:2007ur}
  G.~Bozzi, B.~Jager, C.~Oleari and D.~Zeppenfeld,
  Phys.\ Rev.\  D {\bf 75}, 073004 (2007)
  [arXiv:hep-ph/0701105].

\bibitem{Catani:1996vz}
  S.~Catani and M.~H.~Seymour,
  Nucl.\ Phys.\  B {\bf 485}, 291 (1997)
  [Erratum-ibid.\  B {\bf 510}, 503 (1998)]
  [arXiv:hep-ph/9605323].

\bibitem{Passarino:1978jh}
  G.~Passarino and M.~J.~G.~Veltman,
  Nucl.\ Phys.\  B {\bf 160}, 151 (1979).

\bibitem{Denner:2002ii}
  A.~Denner and S.~Dittmaier,
  Nucl.\ Phys.\  B {\bf 658}, 175 (2003)
  [arXiv:hep-ph/0212259],
  Nucl.\ Phys.\  B {\bf 734}, 62 (2006)
  [arXiv:hep-ph/0509141].

\bibitem{Pumplin:2002vw}
  J.~Pumplin, D.~R.~Stump, J.~Huston, H.~L.~Lai, P.~Nadolsky and W.~K.~Tung,
  JHEP {\bf 0207}, 012 (2002)
  [arXiv:hep-ph/0201195].

\bibitem{kToriginal}
  S.~Catani, Yu.~L.~Dokshitzer, and B.~R.~Webber,
  Phys.\ Lett.\ B  {\bf 285} 291 (1992);
  S.~Catani, Yu.~L. Dokshitzer, M.~H.~Seymour, and B.~R.~Webber,
  Nucl.\ Phys.\  {\bf B406} 187 (1993);
  S.~D.~Ellis and D.~E.~Soper, Phys.\ Rev.\ D
  {\bf 48} 3160 (1993).

\bibitem{kTrunII}
  G.~C.~Blazey {\it et al.},
  arXiv:hep-ex/0005012.
  
\end{thebibliography}
\end{document}